\documentclass{article}

\usepackage{arxiv}
\usepackage{changepage}

\usepackage[font=sf]{caption} % Changes font of captions.

\usepackage{subcaption}
\usepackage{float}
\usepackage{amsmath}
\usepackage{amsfonts} 
\usepackage{amssymb}
\usepackage{siunitx}
\usepackage{verbatim}
\usepackage{hyperref} % Required for inserting clickable links.
\usepackage[dvipsnames]{xcolor}
\usepackage[normalem]{ulem}
\usepackage{dirtytalk}
\usepackage{tikz}

\usepackage[utf8]{inputenc} % allow utf-8 input
\usepackage[T1]{fontenc}    % use 8-bit T1 fonts
\usepackage{url}            % simple URL typesetting
\usepackage{booktabs}       % professional-quality tables
\usepackage{nicefrac}       % compact symbols for 1/2, etc.
\usepackage{microtype}      % microtypography
\usepackage{lipsum}
\usepackage{graphicx}
\graphicspath{ {./images/} }
\usepackage{indentfirst}

\usepackage[backend=biber]{biblatex}
\begin{document}
    
\title{Defunding Sexual Healthcare: A Topological Investigation of Resource Accessibility}

\author{
 Denise Gonzalez-Cruz \\
  Rutgers University-New Brunswick\\
  New Brunswick, NJ 08901 \\
  \texttt{denise.gonzalezcruz@rutgers.edu} \\
  %% examples of more authors
   \And
 Genesis Encarnacion \\
  Mount Mary University\\
  Milwaukee, WI 53222 \\
  \texttt{encarnag@mtmary.edu} \\
  \And
 Kaili Martinez-Beasley \\
  Bryn Mawr College\\
  Bryn Mawr, PA 19010 \\
  \texttt{kmartinezb@brynmawr.edu} \\
   \And
   Robin Wilson \\
   Loyola Marymount University \\
   Los Angeles, CA 90045 \\
   \texttt{Robin.Wilson@lmu.edu} \\
   \And
   Nicholas Arosemena \\
   Brown University \\
   Providence RI 02912 \\
   \texttt{nicholas\_arosemena@brown.edu} \\
   \And
   Atilio Barreda \\
   CUNY Graduate Center \\
   New York, NY 10016 \\
   \texttt{abarreda@gradcenter.cuny.edu} \\
   \And
   Omayra Ortega \\
   Sonoma State University  \\
   Rohnert Park, CA 94928 \\
   \texttt{omayra.ortega@sonoma.edu} \\
   \And
   Daniel A. Cruz \\
   California Polytechnic State University, SLO \\
    San Luis Obispo, California 93407 \\
   \texttt{dcruz66@calpoly.edu} \\
  %% \And
  %% Coauthor \\
  %% Affiliation \\
  %% Address \\
  %% \texttt{email} \\
  %% \And
  %% Coauthor \\
  %% Affiliation \\
  %% Address \\
  %% \texttt{email} \\
}

\maketitle
\begin{abstract}
Government actions, such as the \textit{Medina v. Planned Parenthood South Atlantic} Supreme Court ruling and the passage of the Big Beautiful Bill Act, have aimed to restrict or prohibit Medicaid funding for Planned Parenthood Healthcare Centers (PPHCs) at both the state and national levels. These funding cuts are particularly harmful in states like California, which has a large population of Medicaid users. This analysis focuses on the distribution of Planned Parenthood clinics and Federally Qualified Health Centers (FQHCs), which offer essential reproductive healthcare services including, but not limited to, abortions, birth control, HIV services, pregnancy testing and planning, STD testing and treatment, and cancer screenings. While expanded funding for FQHCs has been proposed as a solution, it fails to address the locational accessibility of Medicaid-funded health centers that provide sexual and reproductive care. To assess this issue, we analyze the proximity of data points representing California’s PPHC and FQHC locations. Topological Data Analysis (TDA)–an approach that examines the shape and structure of data– is used to detect disparities in reproductive and sexual healthcare coverage. To conduct data collection and visualization, we utilize  R and Python. We apply an $n$-closest neighbor algorithm to examine distances between facilities and assess changes in travel time required to reach healthcare sites. We apply persistent homology to analyze current gaps across multiple scales in healthcare coverage and compare them to potential future gaps. Our findings aim to identify areas where access to care is most vulnerable and demonstrate how TDA can be used to analyze spatial inequalities in public health.
\end{abstract}

% keywords can be removed

\section{Introduction}
\label{sec:intro}
\indent The Planned Parenthood Federation of America (PPFA) is a nonprofit organization committed to providing high-quality, inclusive, and comprehensive sexual and reproductive healthcare services to all individuals, regardless of their background or circumstances. These services include, but are not limited to, abortions, birth control, HIV services, pregnancy testing and planning, STD testing and treatment, and cancer screenings \cite{PlannedParenthood}. However, a recent Supreme Court case, Medina v. Planned Parenthood South Atlantic \cite{Medina,medina2025}, along with the passage of a major federal statute popularly known as the One Big Beautiful Bill Act (OBBBA), threatens to significantly reduce PPFA’s revenue by federally defunding the organization. This would place approximately 200 clinics across 24 states at risk of closure \cite{PPdefundedblog2025}. In 2015, $43\%$ of PPFA's revenue was derived from reimbursements for government health service programs, including Medicaid-covered medical services \cite{crsPlannedParenthood}. Previous attempts to pass legislation to federally defund the PPFA turned to Community Healther Centers, including Federally Qualified Health Centers (FQHCs)\textemdash reallocating federal funds initially given to Planned Parenthood Health Centers (PPHCs) to FQHCs\textemdash as possible alternatives to meet the resulting increase in demand \cite{crsPlannedParenthood}.

By law, Medicaid state programs have to cover the services provided by FQHCs to Medicaid beneficiaries \cite{crsPlannedParenthood}. FQHCs are then obligated to grant services to all individuals in need in their service area regardless of their ability to pay, and they must be located in Medically Undeserved Areas (MUAs) \cite{crsPlannedParenthood}. These requisites are established by the Public Health Service Act Section 330 grant funding given to healthcare centers when designated as FQHCs\cite{CRS2}. Compared to other health centers that issue care to Medicaid beneficiaries, including PPHCs, FQHCs are granted higher Medicaid reimbursement rates. Along with previously stated obligations, health centers under Section 330 must deliver preventative health services such as family planning. They also need to be able to refer patients to other providers when the facility can not provide certain services \cite{CRS2}. There are other facilities called FQHC look-alikes that receive the same reimbursement rates, serve in MUAs, and provide universal care; however, these facilities are not governed by the Section 330 grant. Therefore, we did not consider FQHC look-alikes in this report.

In hindsight, the allocation and increase of funds to FQHCs seems like a viable solution to the loss of Planned Parenthood's care. Yet, the expansion of financial resources and thus provided services does not guarantee the client's ability to reach a FQHC. The exclusion of Planned Parenthood from federal funding limits Medicaid-covered patients' choice of provider \textit{and} their locational accessibility to healthcare centers. In this study, we used Topological Data Analysis (TDA) to identify and examine gaps in current and future healthcare resource coverage in California for Medicaid users with a focus on spatial disparities.

Using Persistent Homology (PH), a key tool in TDA, we study the spatial distribution of healthcare centers to identify and quantify gaps in coverage. Specifically, we use PH to detect clusters and holes in our dataset and to track the persistence of these features across different spatial scales \cite{kemme2025persistenthomologypedagogicalintroduction}. Our analysis was done on a point cloud $X=\{x_{i}\}_{i=1}^n$ of health care locations, where $n$ is the total number of facilities. To properly analyze inaccessibility to PPHCs caused by federal defunding, we examined two variations of the point cloud: one in which $x_i$ represents both PPHCs and FQHCs and another in which $x_i$ just represents FQHCs. Both point clouds exist in a metric space $(M,d)$, where $M=\mathbb{R}^2$ and $d$ is a non-Euclidean distance function, which we define in Section \ref{sec:methods}. Given our point cloud $X$, we can compared the distance between points, calculated by $d$, in relation to a scale parameter $r>0$. As $r$ grows, topological features like holes may appear (birth of the hole) and eventually disappear (death of the hole). Holes that persist over a wide range of $r$ are interpreted as significant gaps in healthcare coverage \cite{hickok2023persistenthomologyresourcecoverage}.

Our focus for this study will be in California due to its large population and the high percentage of residents enrolled in Medicaid, which makes access to low-cost reproductive healthcare services critical \cite{StateMedicaid}. We obtained travel times between sites using the Google Maps application programming interface (API). To account for varying modes of transportation and access to them, we collected county-level car ownership data from the California Department of Motor Vehicles (DMV). Travel time acts as the distance metric for the Vietoris-Rips filtration developed in Section \ref{sec:methods}, allowing us to study accessibility in terms of travel burden rather than geographic proximity alone.

\subsection{Background}
Before addressing other research methodology, it is important that we go over TDA and PH, as both are minimally explored forms of analysis for resource coverage. The inspiration for this mode of analysis, and the overall structure of this study, comes from recent work on applying persistent homology to resource coverage \cite{hickok2023persistenthomologyresourcecoverage}.

\begin{figure}[H]
    \centering
    \includegraphics[width=\linewidth]{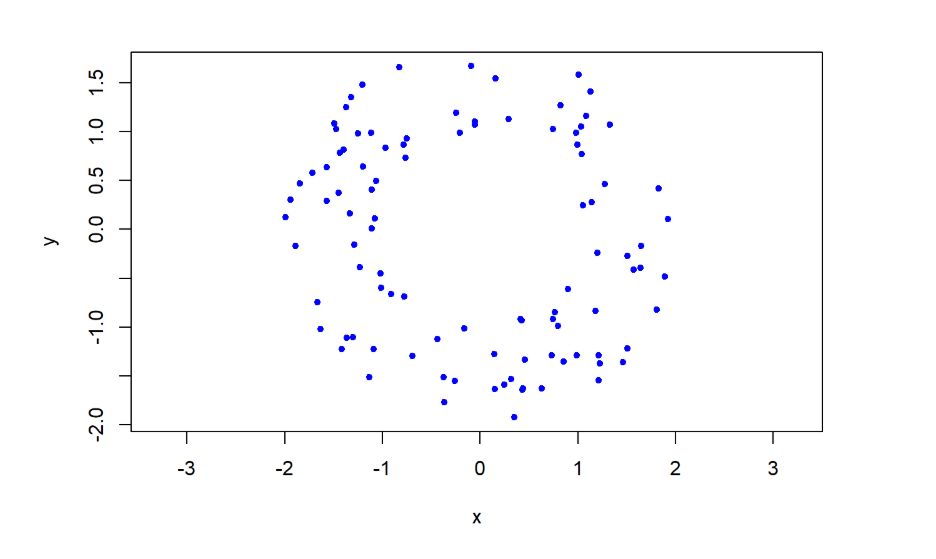}
    \caption{Arbitrary point cloud $W$ with points $w_i$.} \label{arbitrarypointcloud}
\end{figure}

 \textit{Topological data analysis} is an application of topology to data science, so it studies the structural properties of data such as holes\cite{connectingdots}. The data is set within a metric space $M$ defined by a \textit{point cloud} $W$ composed by a finite set of points $w_i$ demonstrative of the data and an associated metric \textit{distance} $d$. A clear example of a hole in a point cloud can be seen in Figure \ref{arbitrarypointcloud}.

\begin{figure}[htb]
    \centering
    \begin{subfigure}[t]{0.4\linewidth}
        \centering
        \includegraphics[width=\linewidth]{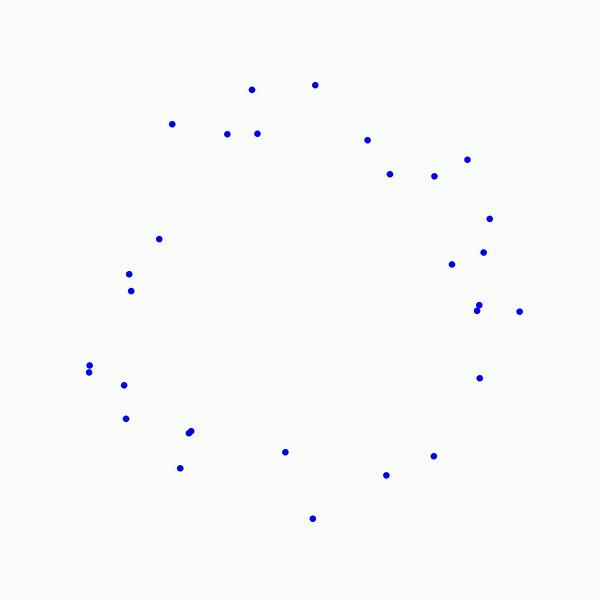}
        \caption{$r_a$}
    \end{subfigure}
    \begin{subfigure}[t]{0.4\linewidth}
        \centering
        \includegraphics[width=\linewidth]{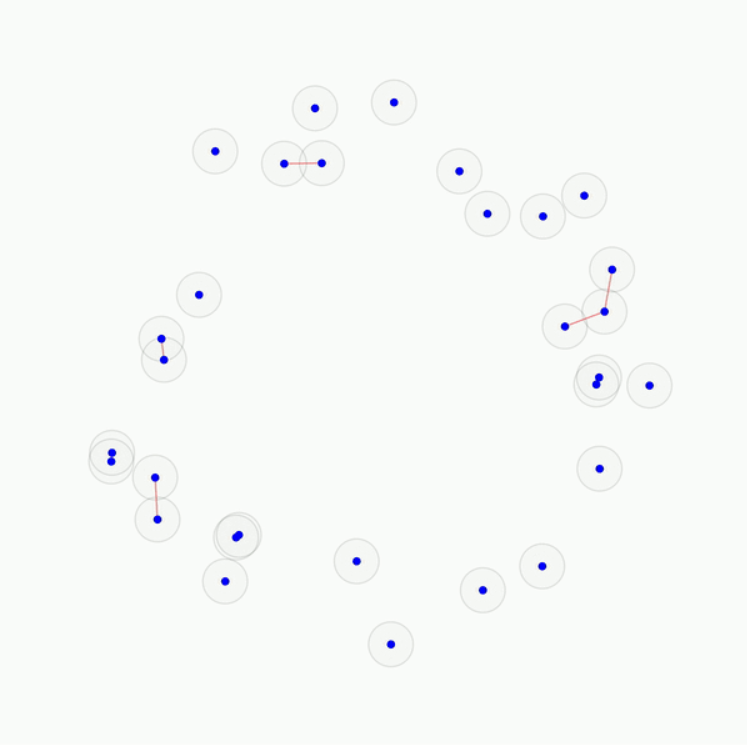}
        \caption{$r_b$}
    \end{subfigure}
    \begin{subfigure}[t]{0.4\linewidth}
        \centering
        \includegraphics[width=\linewidth]{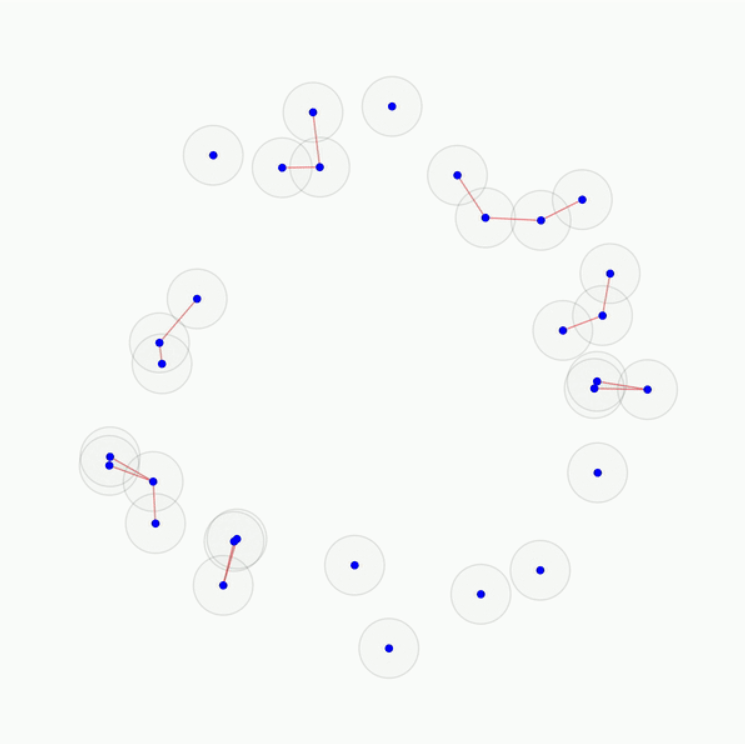}
        \caption{$r_c$}
    \end{subfigure}
    \begin{subfigure}[t]{0.4\linewidth}
        \centering
        \includegraphics[width=\linewidth]{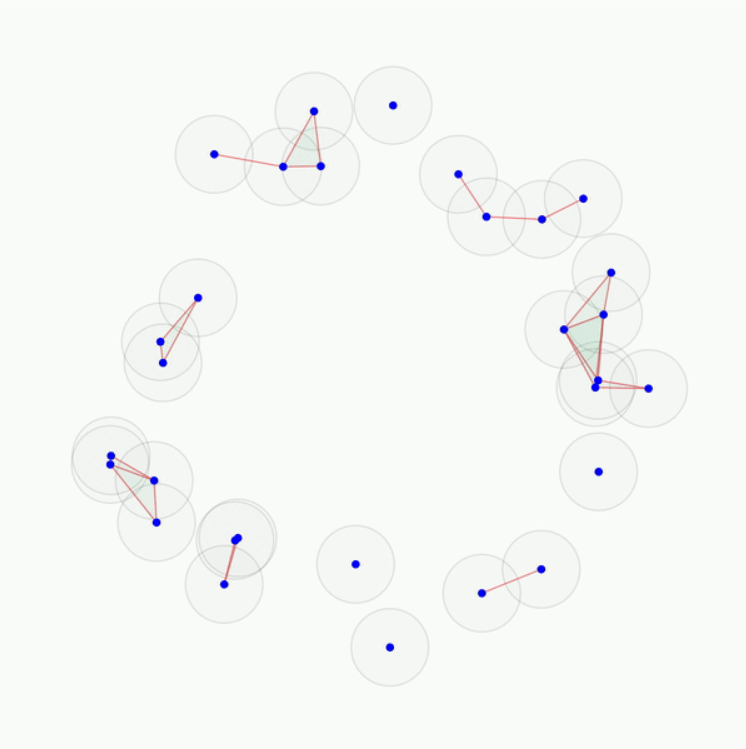}
        \caption{$r_d$}
    \end{subfigure}
    \caption{As the radii ($r$) of each point intersects\textemdash when the distance between each point is less than $2r$\textemdash simplices form between the points. Here, edges emerge in Figures 2.c. and 2.d. \cite{simplicesgif}.}
    \label{fig:filtrationsimplices}
\end{figure}

 In order to apply PH to a point cloud, we must define/identify \textit{simplices}. Because we are analyzing two-dimensional data\textemdash FQHCs and PPHCs ordered by their latitude and longitude\textemdash the simplices we focused on are $0$-dimensional, $1$-dimensional, and $2$-dimensional (a vertex/point, edge, and triangle respectively). A \textit{simplicial complex} is a collection of simplices with the requirement that the \textit{faces} of all simplices are also in the collection. If there is an $n$-dimensional simplex, then its face is the $(n-1)-dimensional$ simplex. Topological data analysis is not just used to view a stagnant structure of data but to determine how the structure of data can change over the increments of a scale. That is when we turn to \textit{filtrations}, which are nested collections of simplicial complexes. A filtration is associated with a \textit{filtration parameter} $r$ that acts as the scale. The filtration starts with an initial simplicial complex and as the filtration parameter increases, more simplices are added and expand upon the previous complex. Each simplicial complex in the filtration is referred to by the scale parameter at which it was formed.  Figure \ref{fig:filtrationsimplices} visualizes the progression of a filtration along the parameter $r$. For this study, we use \textit{Vietoris-Rips (VR) Complexes} and the \textit{VR filtration} that they define.

A VR complex is a complex within a metric space such that a simplex emerges if for any two vertices $x$ and $y$, $d(x,y)<2r$. So, the distance between any two points in the complex must be less than double the scale parameter for a full simplex to form. For the VR complex, the parameter $r$ refers to the radius surrounding the points. Within the VR filtration, the radius around the points increases according to the step size of the parameter, and simplices develop to form new complexes as the distance between vertices/points satisfies $d(x,y)<2r$. We can tell that the inequality has been satisfied visually by the overlapping of the points' radii. In Figure \ref{fig:filtrationsimplices}, we see edges appear in $r_c$ when the radii overlap. In $r_d$, a triangle simplex forms when the radii of all three points overlap.

We apply \textit{homology} to our filtrations to describe the holes that appear in our data. For our research, we limited our analysis to the $0$-dimensional and $1$-dimensional homologies, relating to the n-dimensional simplicies we previously defined. The data's homology can be broken down into two specific components: \textit{homology class} and \textit{homology group}. For $0$D homology, the homology group describes the connectivity between vertices/points of data. This property of connectivity relates to the presence of holes in the data, as space between vertices, ``filled'' by connecting edges, are the holes in the $0$D homology. Each of these holes are associated with a $0$D homology class. The $0$D homology classes are the $1$\!D holes in our data\textemdash they are $1$\!D holes because they are filled by $1$\!D simplicies (edges connecting the vertices). For $1$\!D homology, the homology group detects $2$D holes in the data. The $2$D holes are associated with a $1$\!D homology class, and they are $2$D because they are filled by $2$D simplicies (triangles).

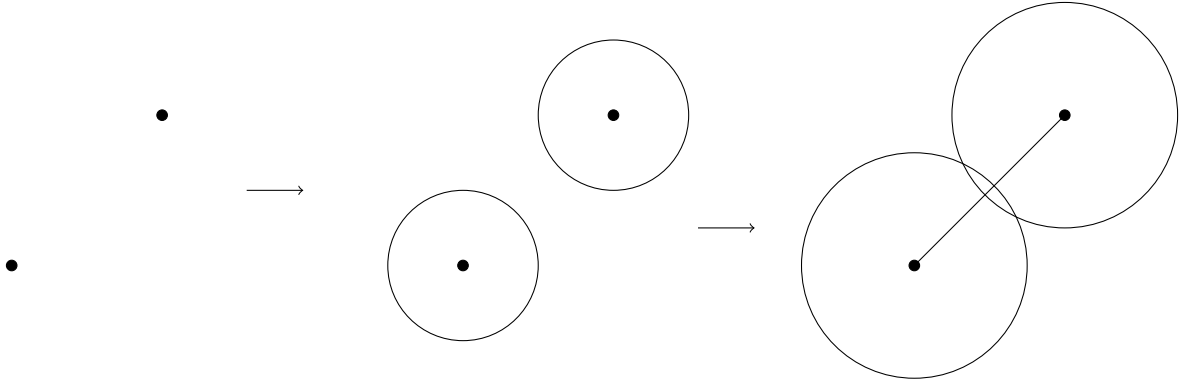
\begin{figure}[H]
\centering

\begin{tikzpicture}
 \filldraw[black] (1,1) circle (2pt);
 \filldraw[black] (3,3) circle (2pt);

\node[draw=none] (invisiblenode1) at (4,2) {};
\node[draw=none] (invisiblenode2) at (5,2) {};
\draw[->] (invisiblenode1) -- (invisiblenode2);

\filldraw[black] (7,1) circle (2pt);
\filldraw[black] (9,3) circle (2pt);
\draw (7,1) circle (1cm);
\draw (9,3) circle (1cm);

\filldraw[black] (13,1) circle (2pt);
\filldraw[black] (15,3) circle (2pt);
\draw (13,1) circle (1.5cm);
\draw (15,3) circle (1.5cm);

\node[draw=none] (invisiblenode3) at (10,1.5) {};
\node[draw=none] (invisiblenode4) at (11,1.5) {};
\draw[->] (invisiblenode3) -- (invisiblenode4);

\draw (13,1) -- (15,3);

\end{tikzpicture}

\caption{A VR filtration through 3 phases. Phases 1 through 3, from left to right, ranges on a scale parameter $r_i$ from $r_0$ to $r_2$. At the $0$D homology, the two vertices are born at $r_0$. There is a homology class in $r_0$ and $r_1$ as the surrounding radii have not yet intersected and there is space between the vertices. In $r_2$, the homology class dies with the edge connecting the two vertices. Therefore, the connecting edge is the death simplex of the homology class and the death value of the homology class is $r_2$.}
\label{fig:0Dhomology}
\end{figure}

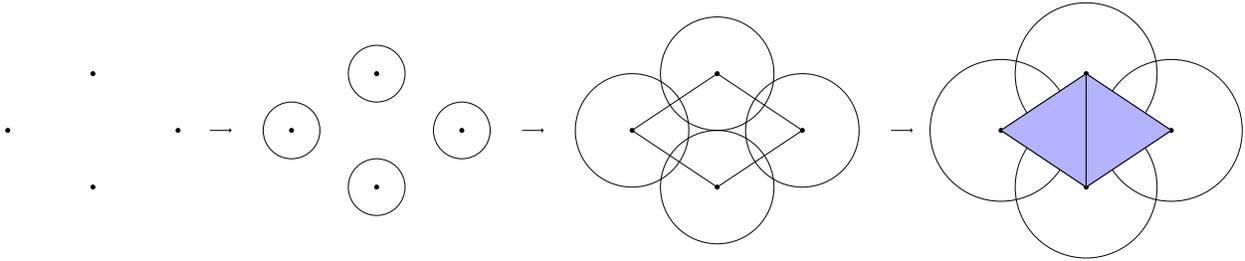
\begin{figure}[H]
\centering

\resizebox{\textwidth}{!}{
\begin{tikzpicture}
 \filldraw[black] (-1,2) circle (2pt);
 \filldraw[black] (2,4) circle (2pt);
  \filldraw[black] (5,2) circle (2pt);
 \filldraw[black] (2,0) circle (2pt);

\node[draw=none] (invisiblenode5) at (6,2) {};
\node[draw=none] (invisiblenode6) at (7,2) {};
\draw[->] (invisiblenode5) -- (invisiblenode6);

\filldraw[black] (15,2) circle (2pt);
\filldraw[black] (12,4) circle (2pt);
\filldraw[black] (9,2) circle (2pt);
\filldraw[black] (12,0) circle (2pt);
 
\draw (15,2) circle (1cm);
\draw (12,4) circle (1cm);
\draw (9,2) circle (1cm);
\draw (12,0) circle (1cm);

\node[draw=none] (invisiblenode7) at (17,2) {};
\node[draw=none] (invisiblenode8) at (18,2) {};
\draw[->] (invisiblenode7) -- (invisiblenode8);

 \filldraw[black] (27,2) circle (2pt);
 \filldraw[black] (24,4) circle (2pt);
  \filldraw[black] (21,2) circle (2pt);
 \filldraw[black] (24,0) circle (2pt);
 
\draw (27,2) circle (2cm);
\draw (24,4) circle (2 cm);
\draw (21,2) circle (2 cm);
\draw (24,0) circle (2 cm);

\draw (24,4) -- (27,2);
\draw (24,4) -- (21,2);
\draw (24,0) -- (27,2);
\draw (24,0) -- (21,2);

\node[draw=none] (invisiblenode9) at (30,2) {};
\node[draw=none] (invisiblenode10) at (31,2) {};
\draw[->] (invisiblenode9) -- (invisiblenode10);

\filldraw[black] (40,2) circle (2pt);
\filldraw[black] (37,4) circle (2pt);
\filldraw[black] (34,2) circle (2pt);
\filldraw[black] (37,0) circle (2pt);
 
\draw (40,2) circle (2.5 cm);
\draw (37,4) circle (2.5 cm);
\draw (34,2) circle (2.5 cm);
\draw (37,0) circle (2.5 cm);

\draw (37,4) -- (40,2);
\draw (37,4) -- (34,2);
\draw (37,0) -- (40,2);
\draw (37,0) -- (34,2);
\draw (37,0) -- (37,4);

\filldraw[fill=blue!30, draw=black, thick] (37,4) -- (34,2) -- (37,0) -- cycle;
\filldraw[fill=blue!30, draw=black, thick] (37,4) -- (40,2) -- (37,0) -- cycle;

\end{tikzpicture}
}

\caption{A VR filtration through 4 phases. Phases 1 through 4, from left to right, range on a scale parameter $r_i$ from $r_0$ to $r_3$. At the $1$\!D homology, the homology class is born at the third phase, $r_3$. There is a homology class in $r_0$ and $r_1$ as the surrounding radii have not yet intersected and there is space between the vertices. In $r_2$, the homology class dies with the edge connecting the two vertices. Therefore, the connecting edge is the death simplex of the homology class and the death value of the homology class in $r_2$.}
\label{fig:1Dhomology}
\end{figure}

 Persistence homology allows us to track the ``birth'' (formation) and ``death'' (disappearance) of homology classes moving across the VR filtration. As the parameter, radius, moves up the filtration's scale and news simplicies are formed, we observe changes to the homology classes. The value of parameter $r$ at which a homology class appears is the \textit{birth value, $r_b$}. For the $0$D homology, each vertex shares a birth value of $0$ because they immediately exist in the metric space at radius $0$. For the $1$\!D homology, the birth value will vary as simplicies emerge and develop enclosed, empty spaces. The value of parameter $r$ at which a homology class disappears is the \textit{death value, $r_d$}. At each death value, there is a corresponding \textit{death simplex}. For example, an edge that fills a $1$\!D homology class is a $0$D death simplex and a triangle that fills a $2$D homology class is a $1$\!D death simplex. 
Figures \ref{fig:0Dhomology} and Figure \ref{fig:1Dhomology} portray the birth and death of homology classes for the $0$-dimensional and $1$-dimensional homologies respectively. To determine the lifetime of a homology class, we compute $r_d-r_b$. A persistence diagram (PD) plots each homology class by $(r_b,r_d)$ in the Euclidean plane.

%-------------------------------

The use of these techniques and tools to examine social issues like resource coverage is modern but not novel. Along with the Hickok's paper that employs topological data analysis and persistent homology to analyze the resource coverage of polling sites, the application of TDA and/or PH has been used to see variations in demographic groups and division by political affiliation. Jakini Auset Kauba uses both mechanisms to inspect racial demographics across cities in the United States \cite{racialsegregation}. Persistence homology can provide deeper insights on the impact of gerrymandering as explored by Moon Duchin\cite{gerrymandering}. This paper's use of TDA to reflect upon sexual healthcare coverage continues a trend of fixing mathematical methods like TDA and PH towards goals of social investigation.

\subsection{Paper Organization}
In Section \ref{sec:methods}, we describe our data retrieval methods and other methodology including our application of PH. We share our results and analysis in Section \ref{sec:result}. The report concludes with a discussion of implications, limitations, and potential directions for future work in Section \ref{sec:discussion}. Our code is available at \url{https://github.com/deniseg20/MSRIUP\_TDA\_Healthcare\_Accessibility.git}.

\section{Methodology}
\label{sec:methods}
\subsection{Creating the Vietoris-Rips Complexes}
\begin{figure}[H]
    \centering
    \begin{subfigure}[t]{0.45\linewidth}
        \centering
        \includegraphics[width=\linewidth]{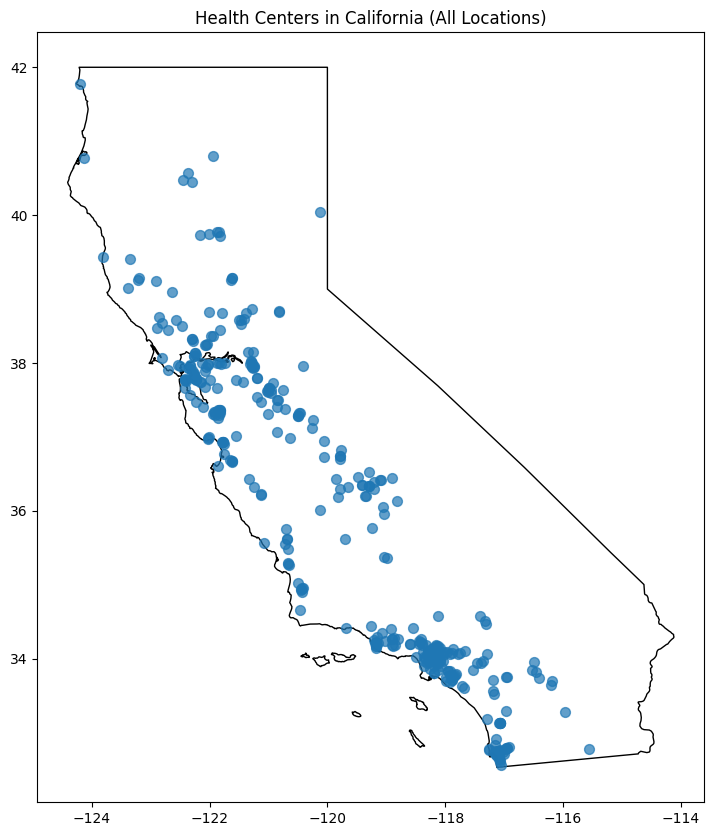}
        \caption{First point cloud variation representing FQHCs and PPHCs}
    \end{subfigure}
    \hfill
    \begin{subfigure}[t]{0.45\linewidth}
        \centering
        \includegraphics[width=\linewidth]{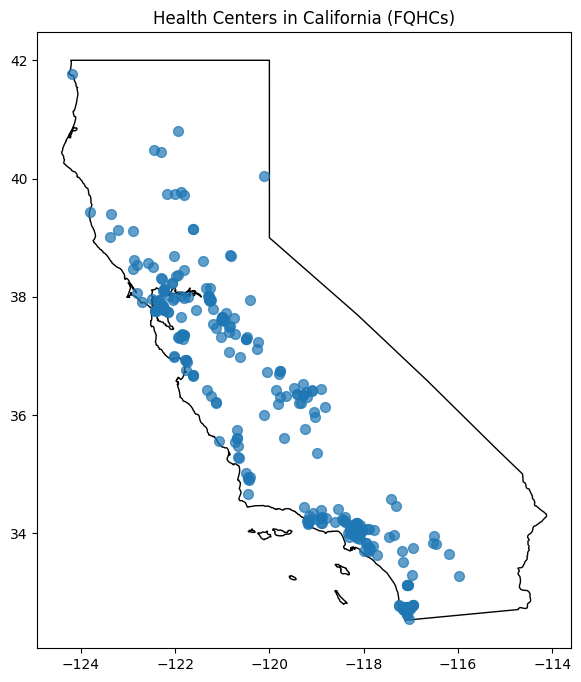}
        \caption{Second point cloud variation representing just FQHCs}
\end{subfigure}
    
    \caption{Point cloud variations for each VR filtration.}
    \label{fig:pointsclouds}
\end{figure}

To fully capture the degrees of coverage with and without the presence of PPHCs, we developed two Vietoris-Rips (VR) filtrations. Each filtration is defined by a point cloud $X={x_i}$ in which each point represents a FQHC and/or a PPHC. In the first VR filtration, each point, or vertex, refers to either a FQHC or PPHC in California. In the second VR filtration, each point or vertex in the point cloud only refers to FQHCs in California. The second filtration is used to analyze a theoretical absence of PPHCs, whether it is a personal absence because Medicaid beneficiaries are no longer covered or the organization dissolving because of extreme revenue cuts. Figures \ref{fig:pointsclouds}(a) and \ref{fig:pointsclouds}(b) visualize the different points included in the filtrations.

While previous uses of TDA and PH used weights to develop their filtrations\textemdash i.e. each point ${x_i}$ is associated with a weight ${w_i}$\textemdash they are not applied to the points in our filtrations. The complex is not developed with a distance metric defined by the Euclidean distance, but rather an estimate of the time it takes to travel to and from one health center to another. Let $x$ and $y$ be two healthcare facilities. We calculate the time to travel from center $x$ to center $y$ as:

\begin{align*}
    \tilde{d}(x,y):= V(C(x))\text{min}\{t_\text{car}(x,y),t_\text{pub}(x,y),t_\text{walk}(x,y)\} \\
    +[1-V(C(x))]\text{min}\{t_\text{pub}(x,y),t_\text{walk}(x,y)\}
\end{align*}

Here, $C(x)$ denotes the county in California where facility $x$ is located, and $V(C(x))$ represents the proportion of cars registered in the county to the county's population. $V(C(x))$ is calculated by dividing the number of vehicles registered in the county by the population in $C(x)$. We used county-level vehicle registration data provided by the California Department of Motor Vehicles (DMV)\cite{californiadmvportal}. The proportion $V(C(x))$ is a way to estimate the proportion of the population in the county who have access to a car. The terms $t_{\text{car}}(x, y), t_{\text{pub}}(x, y)$, and $t_{\text{walk}}(x,y)$ refer to the estimated travel times from $x$ to $y$ by car, public transportation, and walking, respectively. Further details on their computation is provided in Section \ref{TravelTime}. We assumed that those with access to a car will choose to drive, walk, or take public transport depending on which mode of travel is fastest. Therefore, their travel time is $\text{min}\{t_\text{car}(x, y),t_\text{pub}(x, y),t_\text{walk}(x, y)\}$. Conversely, we assumed that those who do not have access to a car, $1-V(C(x))$, will choose to walk or take public transport depending on the fastest mode of travel. Therefore, their travel time is $\text{min}\{t_\text{pub}(x, y),t_\text{walk}(x, y)\}$. This function is used to calculate the travel time from $x$ to $y$ and $y$ to $x$.

Yet, the function $\tilde{d}(x,y)$ is not symmetrical because $\tilde{d}(x,y)\neq{\tilde{d}(y,x)}$. Not only are the counties' populations and vehicle registration different, but the function also can not take factors like one-way roads and changes to rail lines into consideration. To develop the VR filtration, we need a symmetric function. To address this issue, we took a weighted average of $\tilde{d}(x,y)$ and ${\tilde{d}(y,x)}$. The populations of counties $C(x)$ and $C(y)$ act as the weights. The final distance function is thus defined as:
\begin{align*}
     d=\frac{1}{P}[P_{C(x)}\tilde{d}(x,y)+P_{C(y)}\tilde{d}(y,x)]
\end{align*}

\noindent $P_{C(x)}$ and $P_{C(y)}$ are the populations in counties $C(x)$ and $C(y)$, and $P$ is the sum of both populations. The formation of structures within the data, more specifically holes and simplices, now depend on the distance between points $x$ and $y$ and its relation to the scale parameter $r$. Though the distance function does not share the metric property of triangle inequality, it can still be used to develop our filtrations \cite{hickok2023persistenthomologyresourcecoverage}.\par 

\subsection{Estimating of Travel Time}\label{TravelTime}
We use the Google Maps Routes API to estimate travel times between FQHCs and PPHCs by car, public transportation, and walking, generating a distance matrix for each mode of transportation. Due to budget and time constraints, we limited our queries to each facility’s $35$ nearest neighbors. Neighbors are defined by closest geographic proximity. For each mode of transportation, we calculated both the inbound and outbound travel times between healthcare centers. Our analysis is conducted under two cases, one including both PPHCs and FQHCs, and one including only FQHCs.

If we treat location as a vertex, we are left with three different connected and weighted graphs. A connected graph being one that has an existing path, edge, between any two vertices. Each graph represents a mode of transportation, driving, walking, or public transport, and consists of edges weighted by travel times gathered with the Google API. For example, the edges of the connected graph for driving, $D$, are weighted according to travel time by car from a vertex to its $35$ nearest neighbors. The same idea then applies to walking and public transport. The weight of the edge may change depending on the direction of vertices, whether one is going from $x$ to $y$ or $y$ to $x$, because travel times to and from are not immediately identical. Rather than define the direction of the edges separately, we apply one weight under the assumption that the same mode of transportation is taken to and from one health facility to another. We define the travel time by car, public transport, or foot as a sum of the travel time from $x$ to $y$ ($\tilde{t}(x,y)$) and $y$ to $x$ ($\tilde{t}(y,x)$):
\begin{align*}
    t_{\text{car}}:=\tilde{t}_{\text{car}}(x,y)+\tilde{t}_{\text{car}}(y,x),\\
    t_{\text{pub}}:=\tilde{t}_{\text{pub}}(x,y)+\tilde{t}_{\text{pub}}(y,x),\\
    t_{\text{walk}}:=\tilde{t}_{\text{walk}}(x,y)+\tilde{t}_{\text{walk}}(y,x).\\
\end{align*}

\subsection{Facility Locations}
We manually collected the locations of Planned Parenthood clinics from the organization's official website\cite{PPHCLocations}. Furthermore, we used web scraping techniques to obtain the locations of Federally Qualified Health Centers (FQHCs) from the California Department of Health Care Access and Information\cite{californiahcaiportal}, excluding facilities that do not offer comparable reproductive health services and do not receive the same degree of federal funding from Medicaid reimbursements. Because vehicle registration is provided at the county level, we used the facilities respective counties to represent their surrounding area. 

\begin{figure}[H]
    \centering
    \begin{subfigure}[t]{0.45\linewidth}
        \centering
        \includegraphics[width=\linewidth]{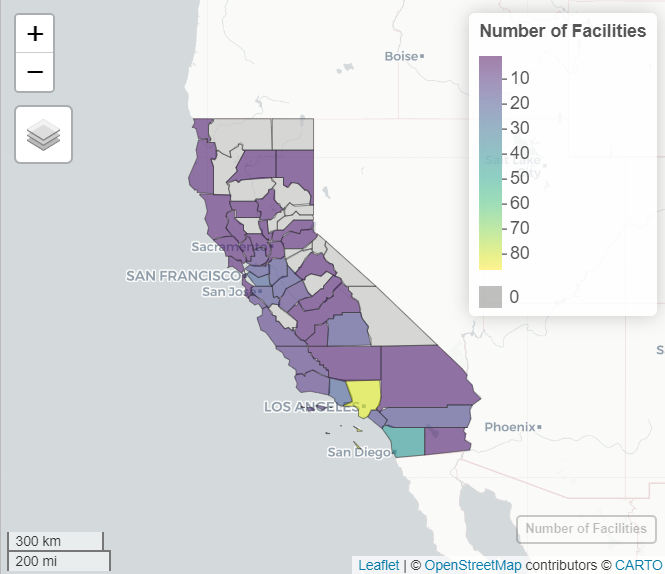}
        \caption{Sum of FQHCs \textit{and} PPHCs by county}
    \end{subfigure}
    \hfill
    \begin{subfigure}[t]{0.45\linewidth}
        \centering
        \includegraphics[width=\linewidth]{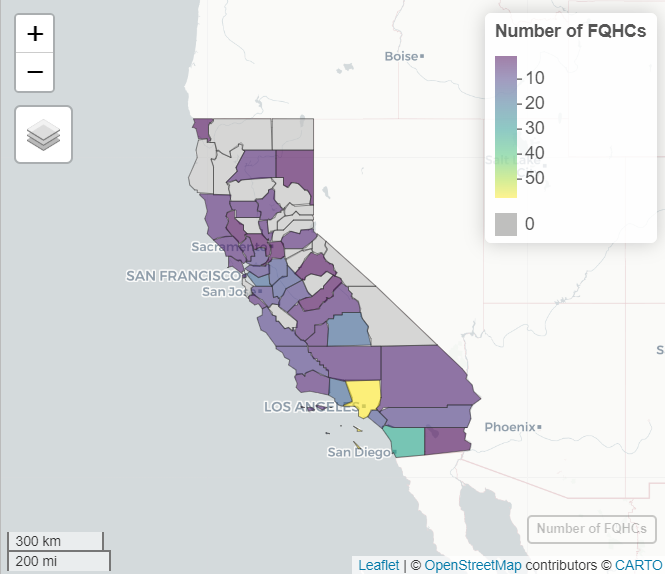}
        \caption{FQHCs by county}
    \end{subfigure}
    \caption{Differences in the heatmaps visualizes the decrease in healthcare facilities that provide sexual healthcare when PPHCs  are removed from California.}
    \label{fig:heatmaps}
\end{figure}

\subsection{Code for the Persistence Diagrams}\label{PDCode}
The code used to construct the Persistence Diagrams (PD) and their associated visuals, which can be seen in Section \ref{sec:result}, and the distance matrix tracking the distance between each location and their neighbors heavily derived from the Jupyter Notebook (\url{https://bitbucket.org/jerryluo8/coveragetda/src/main/}) provided by Hickok et.al \cite{hickok2023persistenthomologyresourcecoverage}. Alterations made to the code to fit this study can be seen in our Github provided in the Paper Organization subsection of Section \ref{sec:intro}. 

\subsection{Significance Testing}\label{Sig.Tests}
For our analysis of $0$D-homology, we employed significance testing to compare connectivity, specifically, travel times between facilities. We considered several statistical tests based on their distribution and variance assumptions. These included the t-test, a test that assumes the data follows a normal distribution and equal variance; the Welch's t-test, which assumes unequal variances and normal distribution; the Mann-Whitney test, which assumes equal variance but does not assume normal distribution; and the Brunner-Munzel test, which, unlike the Mann-Whitney test, assumes unequal variance. We ultimately decided to use the Mann-Whitney test and the Brunner-Munzel test. This is because the datasets failed the normality test, therefore failing one of the requirements for the t-test and Welch's t-test.  First,  the data was trimmed by removing observations with death times of 15 minutes or less to reduce noise. Then, the log-transformation was applied to approximate the data to the normal distribution. We then performed one-tailed versions of the Mann-Whitney test and the Brunner-Munzel test. The results are presented and discussed in Section \ref{sec:result}.

\section{Results}
\label{sec:result}
\begin{figure}[H]
    \centering
    \begin{subfigure}[t]{0.45\linewidth}
        \centering
        \includegraphics[width=\linewidth]{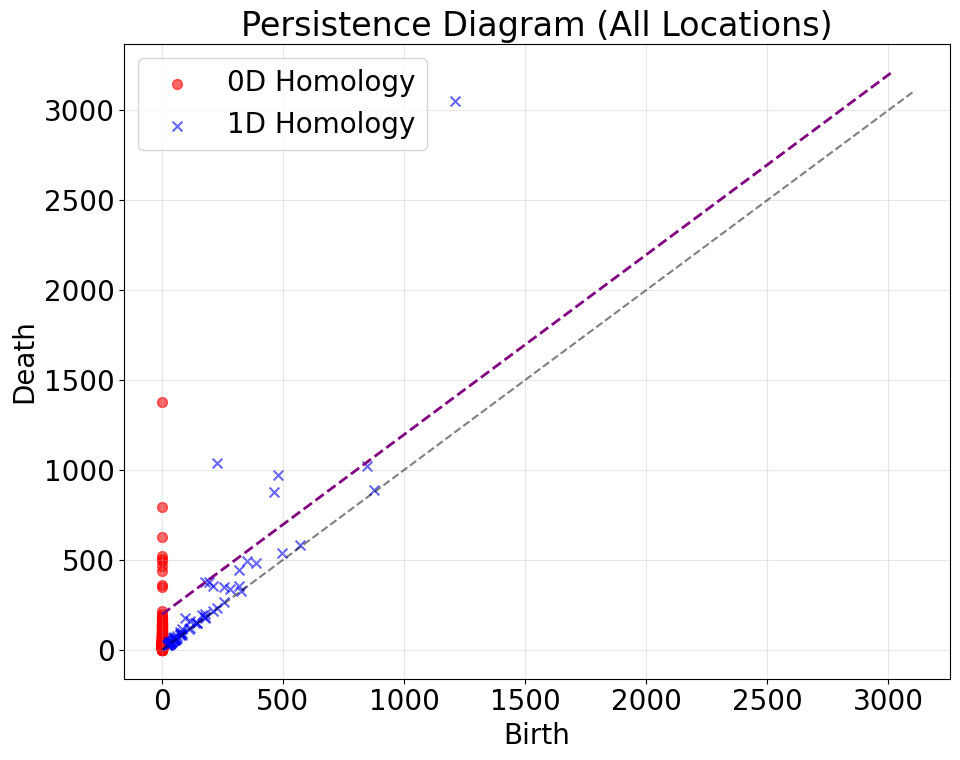}
        \caption{All Locations}
    \end{subfigure}
    \hfill
    \begin{subfigure}[t]{0.45\linewidth}
        \centering
        \includegraphics[width=\linewidth]{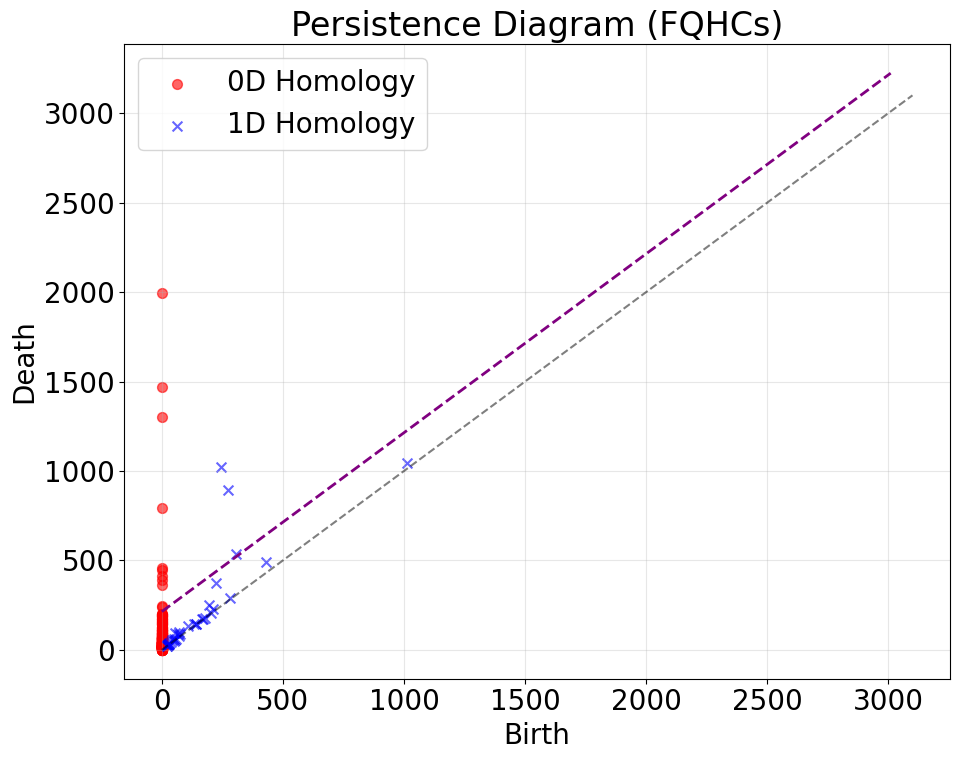}
        \caption{FQHCs}
    \end{subfigure}
    \caption{Persistence Diagrams for PH of VR Complexes}
    \label{fig:PDs}
\end{figure}

After running the PD code, mentioned in Section \ref{PDCode}, we see the resulting images (Fig: \ref{fig:PDs}). The Persistence Diagram (PD) for all health center locations \ref{fig:PDs}(a) shows all points are connected in the $0$-dimensional space, marked by `o', before the points reach 1500 minutes (25 hrs). The white space between the points along the $y$-axis indicates longer travel times to the health centers. The `x' markers represent cycles between more than 2 points in the 1-dimensional space. Highly persistent points appear at or above the mean distance of 196.61 minutes, shown as the purple dotted line. For all locations, there are 5 death simplices at or above the mean distance.                       
  
In the PD for the Federally Qualified Health Centers (FQHC), there's a shift in the death times visible in the $0$-dimensional points after the removal of Planned Parenthood (PP) locations. Comparing it to the PD for all locations, there is a greater distance, 214 minutes, between points before they are all connected. The greatest distance recorded for points to reach connectedness, having an edge between them, is approximately 2000 minutes (about 33 hours) in the $0$D. Observing the $1$\!D space, three simplices persist at or above the mean distance, about 214 minutes.

% Adding in 0D persistence diagrams
\begin{figure}[H]
    \centering
    \begin{subfigure}[t]{0.45\linewidth}
        \centering
        \includegraphics[width=\linewidth]{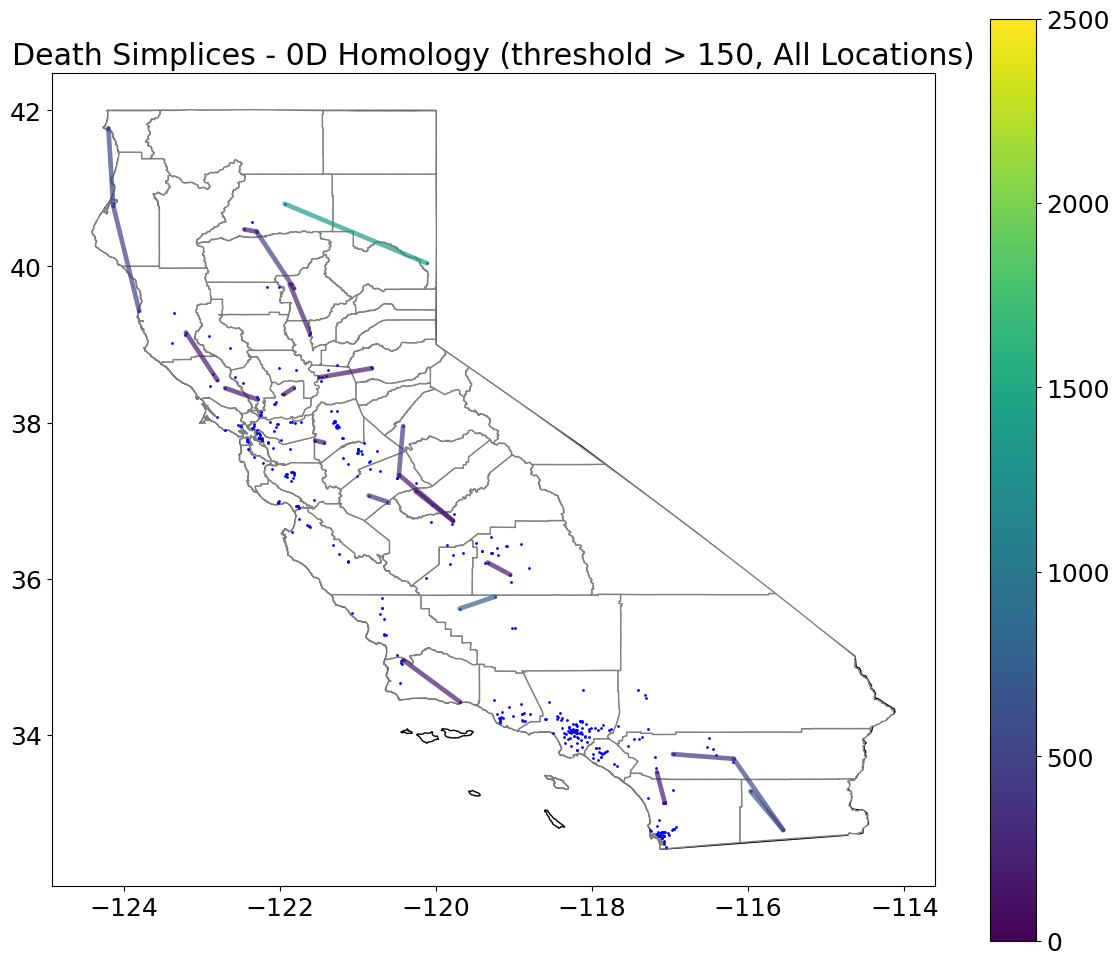}
        \caption{All Locations}
    \end{subfigure}
    \hfill
    \begin{subfigure}[t]{0.45\linewidth}
        \centering
        \includegraphics[width=\linewidth]{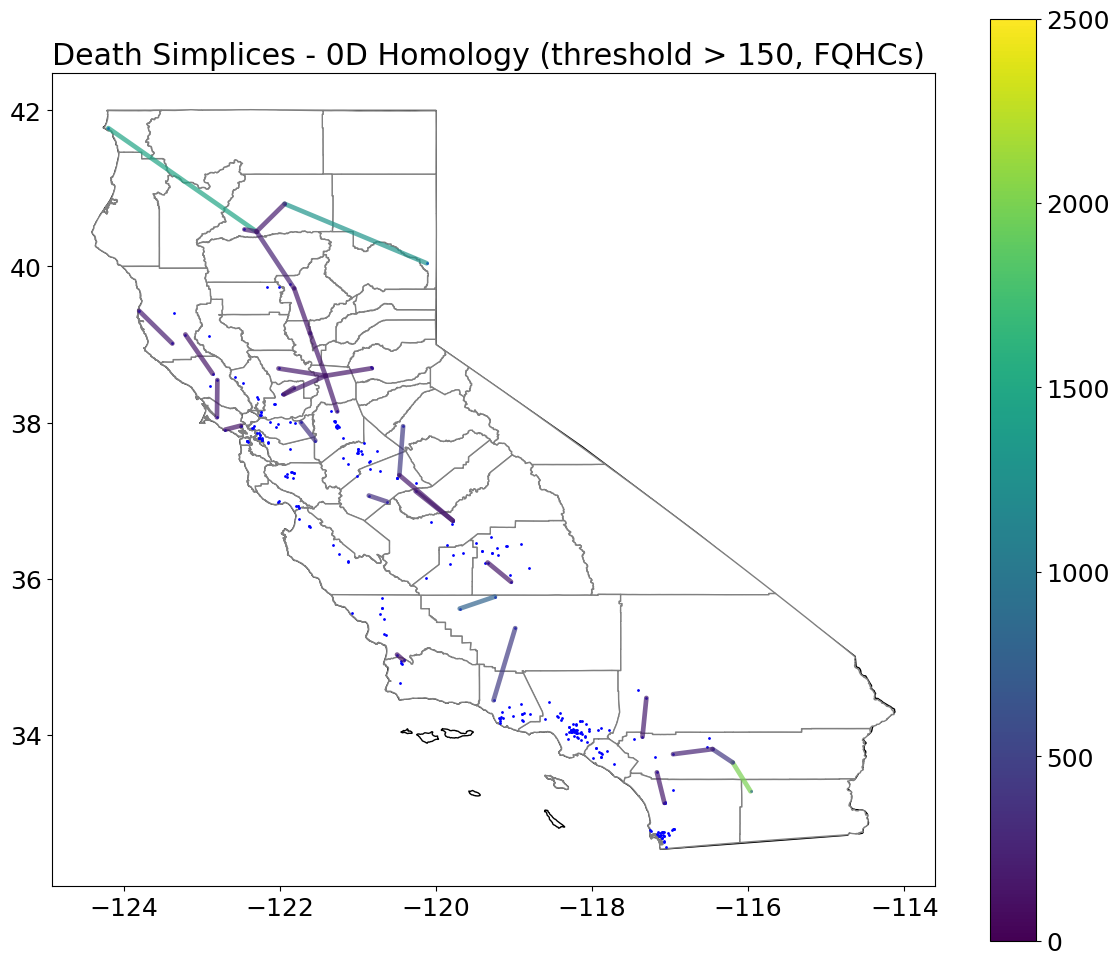}
        \caption{FQHCs}
    \end{subfigure}
    \caption{0D Homology Death Simplices}
    \label{fig:DeathSimplices0D}
\end{figure}

% 1D persistence diagrams
% talking about the death simplices plots

From the PD code, the persistent homology information was used to plot the death simplices in the $0$D and $1$\!D spaces. The death simplices with a distance greater than 150 minutes (2.5 hrs) in $0$D space for all locations and FQHCs (Fig: \ref{fig:DeathSimplices0D}). After removing Planned Parenthood locations, there is an increase in distance in the area between longitudes 40-42 and below longitude 34, where distances are greater than 1000 minutes (approx. 16 hours).

\begin{figure}[H]
    \centering
    \begin{subfigure}[t]{0.45\linewidth}
        \centering
        \includegraphics[width=\linewidth]{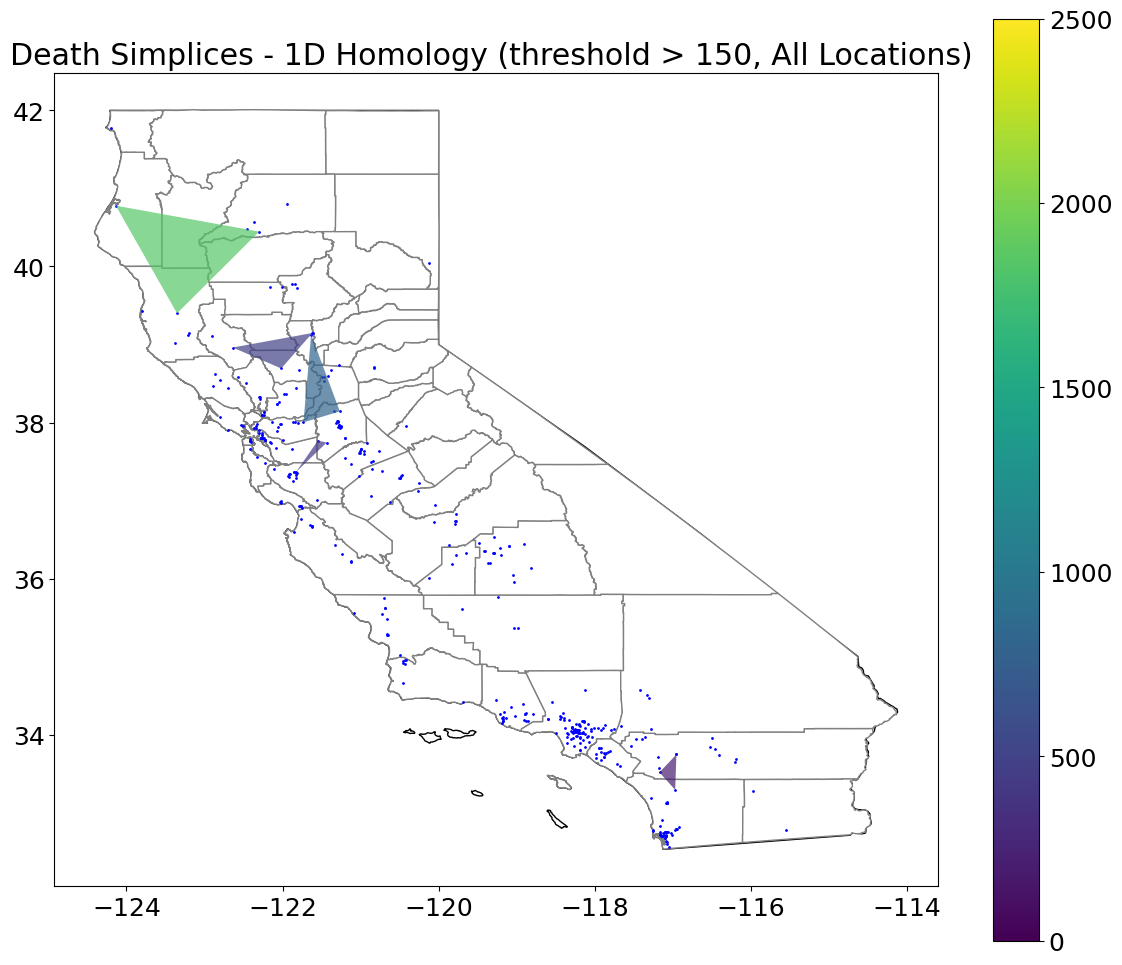}
        \caption{All Locations}
    \end{subfigure}
    \hfill
    \begin{subfigure}[t]{0.45\linewidth}
        \centering
        \includegraphics[width=\linewidth]{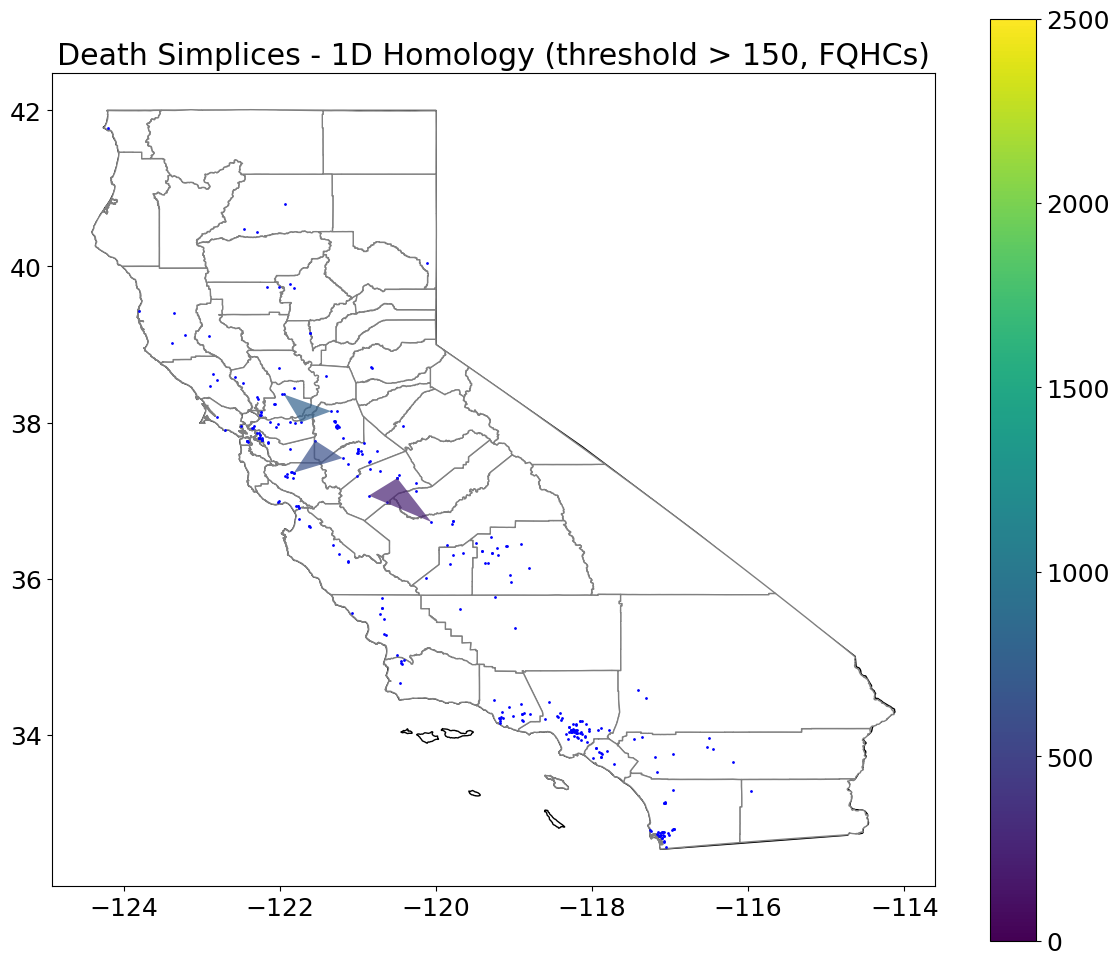}
        \caption{FQHCs}
    \end{subfigure}
    \caption{1D Homology Death Simplices}\label{Fig:DeathSimplices1D}
\end{figure}

Analyzing the graphs for the 1D-Homology, before the removal of Planned Parenthood locations, the 5 death simplices from the PD appear on the map. The largest hole that persists has a distance greater than 1500 minutes. The triangles formed in $1$\!D show the areas that lack access or have greater barriers reaching health centers for care, before the removal of Planned Parenthood locations.

After the removal of Planned Parenthood locations, the largest hole from all locations seems to disappear. This indicates a gain in coverage. However, the mean distance before and after the removal of Planned Parenthood has increased, which is clear when looking at the $0$D Homology death simplices. While not captured in the $1$\!D Homology death simplices, there seems to be a loss of coverage in areas where there previously was access to health centers. Other areas see an increase in already underserved areas. Overall, the removal of Planned Parenthood and/or abortion providing clinics will increase the obstacles faced by underserved communities to access sexual and reproductive healthcare, while creating a lack of access to sexual and reproductive healthcare in new areas.
\begin{figure}[H]
    \centering
    \begin{subfigure}[t]{0.45\linewidth}
        \centering
        \includegraphics[width=\linewidth]{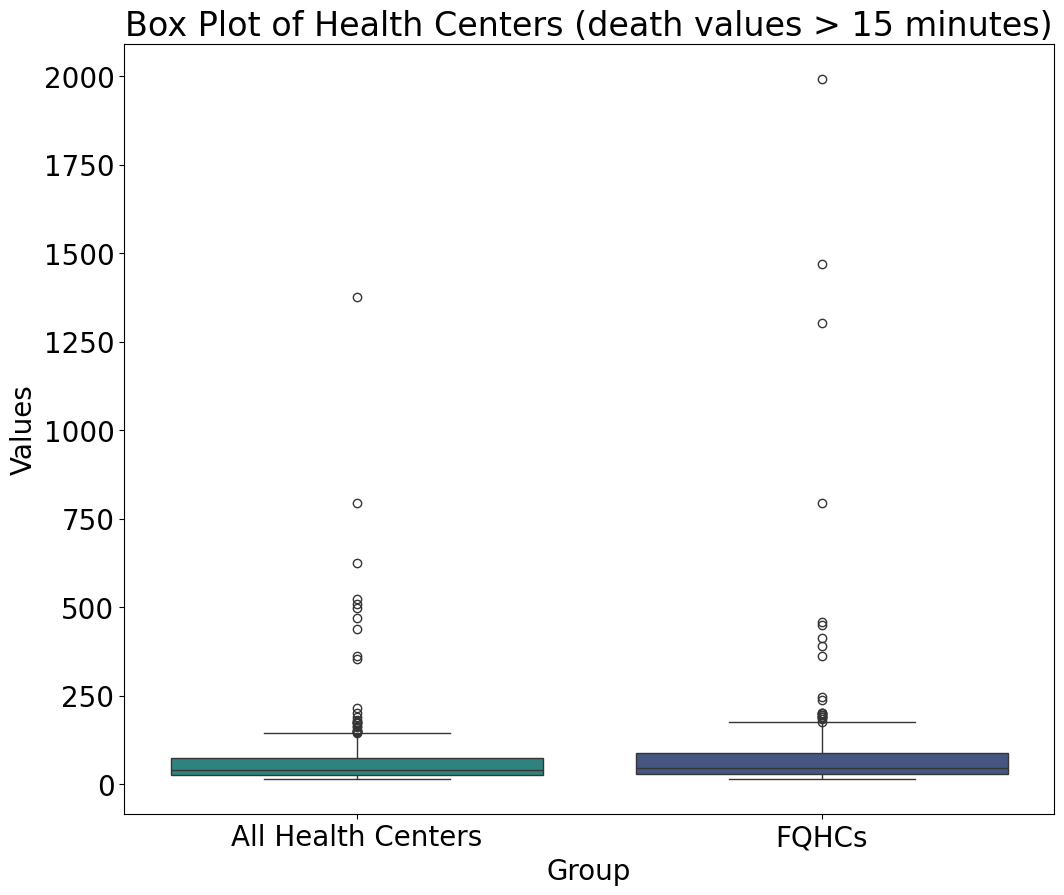}
        \caption{Before applying natural log}
    \end{subfigure}
    \hfill
    \begin{subfigure}[t]{0.45\linewidth}
        \centering
        \includegraphics[width=\linewidth]{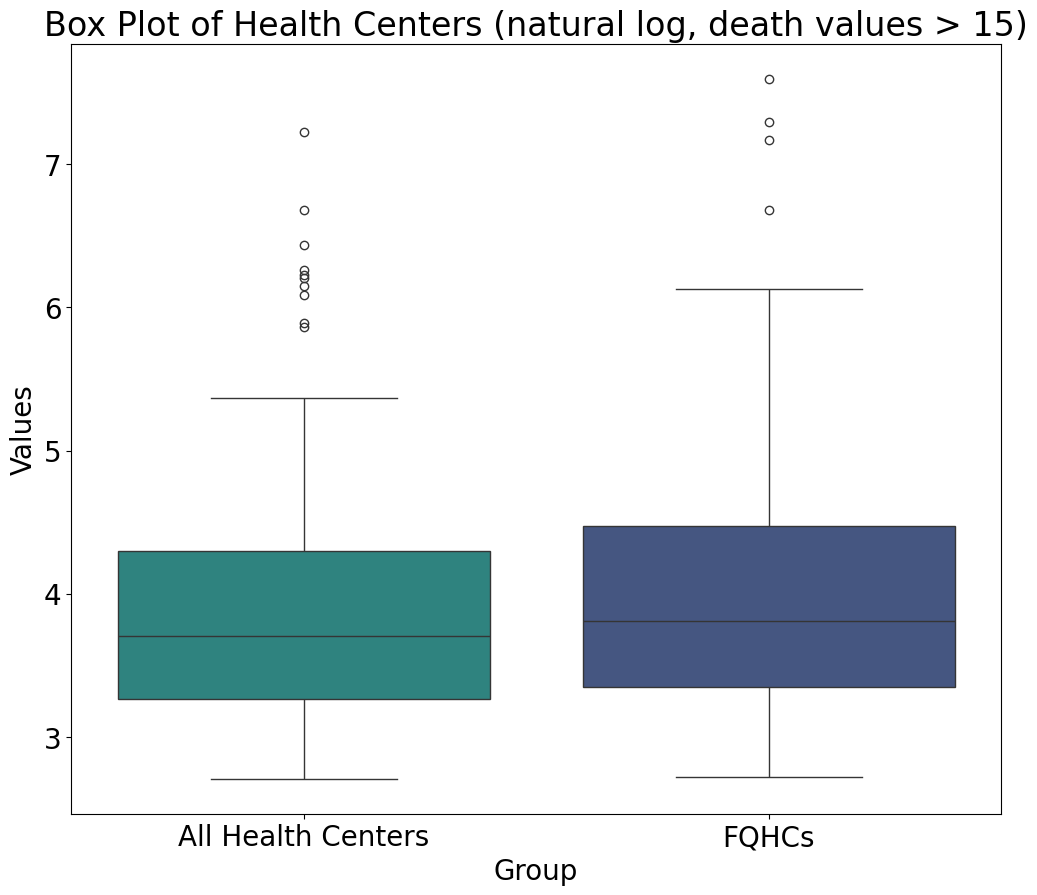}
        \caption{After applying natural log}
    \end{subfigure}
    \caption{Box Plots}\label{Fig:Boxplots}
\end{figure}
    To determine the significance of the changes to health center accessibility before and after the removal of PPHCs, a one-tailed t-test was conducted, as mentioned in Section \ref{sec:methods}.  The null hypothesis, $H_0$,  was that the mean distance to all the health centers, $\mu_a$, was greater than or equal to the mean distance for the FQHCs, $\mu_f$. Where the alternative hypothesis,$H_a$, was $\mu_a$ < $\mu_f$.  The Mann-Whitney test (assuming equal variance) resulted in a p-value of approximately 0.0288, which was statistically significant at 95\% confidence. For the Brunner-Munzel test (assuming unequal variance), the resulting p-value was approximately 0.0291, statistically significant at 95\% confidence. Therefore, we can reject $H_0$ in favor of $H_a$ 
    The results of the tests showed that there is a change in travel time to the health centers after the removal of Planned Parenthood Health Centers.  Therefore, the loss of funding to Planned Parenthood Health Centers
    
\section{Discussion}\label{sec:discussion}

\subsection{Summary}    
We used persistent homology (PH), a method from topological data analysis (TDA), to examine the accessibility of sexual and reproductive healthcare facilities in California. Our analysis focused on travel times, by walking, public transit, and car, to and from various facilities. We considered two scenarios: one including both Federally Qualified Health Centers (FQHCs) and Planned Parenthood Health Centers (PPHCs), and another considering only FQHCs. For each case, we constructed a filtration where a homology class that dies at time $t$ corresponds to a geographic region that requires $t$ minutes of travel to reach a facility and return. We identified the death simplex of a homology class as the location of a potential gap in healthcare coverage. This approach allowed us to evaluate current levels of accessibility and identify existing coverage gaps, as well as predict new gaps that might arise from the potential closure of PPHCs. The persistent diagrams reveal that removing Planned Parenthood locations increases travel distances and delays full connectivity between health centers, especially in already under served areas. While some access gaps in the $1$\!D Homology appear to close, average distances to care rise overall, indicating a broader decline in accessibility. The removal of Planned Parenthood clinics leads to increased travel times, reduced connectivity, and more persistent barriers. Importantly, this method captures a lower bound on the severity of accessibility gaps, since including wait times and financial barriers would only worsen the lack of access.

\subsection{Limitations}\label{sec:limitations}

One limitation of our study is the absence of data on waiting times at Planned Parenthood clinics and Federally Qualified Health Centers (FQHCs). Accurately estimating these times would require access to mobile device Global Positioning System (GPS) ping data at these locations, which is not publicly available. Thus, as discussed in Section \ref{sec:methods}, we utilized an unweighted Vietoris-Rips filtration, which does not account for waiting time variability across facilities. Our analysis therefore focuses only on geographic accessibility, using travel time as the distance metric. Incorporating GPS mobile ping data could improve future analysis by providing more precise estimates of visit duration and providing deeper insight into which facilities deliver more efficient care.

Another limitation of our study is using vehicle registration data in place of car ownership data. We obtained county-level vehicle registration data from the California Department of Motor Vehicles, which provides only a subset of actual car ownership and distribution. This introduces two assumptions, first, that vehicle registration accurately reflects car ownership, and second, that vehicle access is uniformly distributed across all ZIP codes within a county. Due to data availability, we were unable to access car ownership statistics at the ZIP code level. Access to more granular data would reduce the need for these assumptions and allow for a more accurate analysis of transportation access to Planned Parenthood facilities.

Another limitation of our study was the inability to compute travel times and distances between all pairs of points in our datasets. Due to budget constraints, we limited our distance matrix to include only the travel times and distances to the nearest 35 facilities for each location. While this was sufficient for our analysis, having a complete distance matrix, effectively forming a fully connected graph, would have provided a more comprehensive picture of resource coverage across California. Along with not considering all locations within California, we did not examine facilities in the border to states surrounding California. This leads to the assumption, for example, that California residents on the Eastern border of the state will not go to Nevada for healthcare.

\subsection{Future work}\label{sec:future}
A next step for this paper is to focus on the counties where coverage gaps currently exist and identify the ZIP codes within them. This will allow us to use census tracts or other geographic units to gather demographic information and examine whether there is a correlation between these gaps and the presence of minority populations. Through this analysis, we can better understand which groups are currently underserved in California and which are most at risk of losing additional access with the potential loss of PPHCs.

As discussed in Section \ref{sec:limitations}, we chose to work with more readily available county-level data rather than ZIP code-level data. For future analyses of resource availability, ZIP code-level data, such as fuel type distributions, could be cleaned and utilized to provide a more granular understanding. This would also require organizing our PPHC and FQHC data by ZIP code during the analysis.

Another potential refinement for future work involves filtering the FQHC dataset to ensure that the facilities included offer the same types of reproductive health services as PPHCs. This step would help improve the comparability of provider types in our analysis.

An extension of this work would be to scale the analysis from California to the entire United States. In that case, incorporating travel times and distances between all locations would be essential for evaluating national patterns of accessibility. Since access to PPHCs and other reproductive healthcare facilities is especially critical in certain states, this broader scope could offer a more comprehensive understanding of the potential impact of PPHC closures. However, this extension would require manually compiling a complete list of PPHC and FQHC locations across the U.S., which would be time-intensive and beyond the capability of this study. It would also be necessary to compute complete distance matrices, resulting in an immense increase in the number of API queries, therefore a significantly higher cost.

A loss in funding to centers like Planned Parenthood due to varying abortion laws across the US would affect the economy \cite{abortionBanCost}. We can then investigate financial impacts on individuals and the country as a whole.

\section{Acknowledgments}
This material is based on work supported by the National Science Foundation under Grant No. DMS-2149642 and the Sloan Grant under Grant No. G-2024-22394 while the authors participated in a program hosted by the Simons Laufer Mathematical Sciences Institute (formerly Mathematical Sciences Research Institute) in Berkeley, California, during the summer of 2025.

\printbibliography

\end{document}